\documentclass[11pt]{article}
\usepackage{graphics}   
\usepackage{amssymb}
\usepackage{amsmath}
\newtheorem{theorem}{Theorem}

\newtheorem{remark}{Remark}
\usepackage{float}
\floatstyle{ruled}
\newfloat{Algorithm}{tbp}{lop}

\title{Practical Binary Adaptive Block Coder}

\author{
Yuriy A. Reznik \\
QUALCOMM Incorporated \\
5775 Morehouse Drive, San Diego, CA 92121 \\
E-mail: {\tt yreznik@ieee.org \/}}

\date{}

\begin{document}

\maketitle

\begin{abstract}
This paper describes design of a low-complexity algorithm for
adaptive encoding/decoding of binary sequences produced by
memoryless sources. The algorithm implements universal block codes
constructed for a set of contexts identified by the numbers of
non-zero bits in previous bits in a sequence.

We derive a precise formula for asymptotic redundancy of such codes,
which refines previous well-known estimate by Krichevsky and
Trofimov~\cite{KrichevskyTrofimov}, and provide experimental
verification of this result.

In our experimental study we also compare our implementation with
existing binary adaptive encoders, such as JBIG's Q-coder~\cite{QCoder}, and MPEG AVC
(ITU-T H.264)'s CABAC~\cite{CABAC} algorithms.
\end{abstract}

\section{Introduction}
One of the most basic tasks in the design of today's data
compression algorithms is the one of converting input sequences of
bits with some unknown distribution into a decodable bitstream.
This happens, for example, in the design of image or video codecs,
scalable (bit-slice based) encoding of spectrum in audio
codecs, etc. In most such cases, the bits to be encoded are taken from
values produced by various signal processing tools (transforms, prediction
filters, etc), which means that they are already well de-correlated,
and that assumption of memorylessness of such a source is justified.

Most commonly, the problem of encoding of such sequences of bits is
solved by using fast (typically multiplication-free) approximations of
binary adaptive arithmetic codes.
Two well known examples of such algorithms
are IBM's Q-coder~\cite{QCoder} adopted in JBIG image coding
standard~\cite{T82}, and CABAC encoder~\cite{CABAC} used
in MPEG AVC/ITU-T H.264 standards for video compession~\cite{H264}.

In this paper we describe an alternative implementation of adaptive encoder
using an array of Huffman codes designed for several estimated densities,
indexed by the numbers of non-zero bits in previous blocks (contexts)
in a sequence.

We study both efficiency and implementation aspects of such a scheme and
show that by using even relatively short blocks ($8...16$ bits, and using
correspondingly $1.5...5K$ bytes of memory) it can achieve compression
performance comparable or superior to one of the above quoted algorithms.

This paper is organized as follows. In Section 2 we provide background
information about our coding problem. In Section 3 we quote known results about
efficiency of such codes and offer a more precise result.
In Sections 4 and 5 we describe design of our system, and in Section 6 we
provide experimental results. Appendix A contains proofs of out Theorem 1,
and Appendix B contains complete code of the program we've designed.

\section{Background Information}
Consider a memoryless source producing symbols from a binary alphabet
$\left\{0,1\right\}$ with probabilities $p$, and $q=1-p$ correspondingly.
If $w$ is a word of length $n$ produced by this source, then its probability:
\begin{equation}
\Pr\left(w\right) = p^k\, q^{n-k}\,, \label{eq:p_s}
\end{equation}
where $k$ denotes the number of $1$'s in this word (sometimes $k$ is also referred
to as {\em weight\/} of $w$).

A {\em block code\/} $\phi$ is an injective mapping between words
$w$ of length $|w|=n$ and binary sequences (or {\em codewords\/})
$\phi\left(w\right)$:
\begin{equation}
\phi: \left\{0,1\right\}^n \rightarrow \left\{{0,1}\right\}^*\,, \label{eq:phi1}
\end{equation}
where the codewords $\phi\left(w\right)$ represent a {\em uniquely decodable\/} set~\cite{CoverThomas}.

Typically, when the source (i.e. its probability $p$) is known, such a code is designed
to minimize its average length, or (in relative terms) its {\em average redundancy\/}:
\begin{equation}
R_{\phi}\left(n,p\right) = \frac{1}{n} \sum_{|w| = n} \Pr(w)
\left|\phi(w)\right| - H(p)\,. \label{eq:CS}
\end{equation}
As customary by $H(p) = - p \log p - q \log q$ we denote the {\em entropy\/} of
the source~\cite{CoverThomas}.

Classical examples of codes and algorithms suggested for solving
this problem include Huffman~\cite{Huffman}, Shannon~\cite{Shannon},
Shannon-Fano~\cite{Fano}, Gilbert-Moore~\cite{GilbertMoore} codes
and their variants~\cite{Abrahams98}. Performance of such codes is
well studied, see, e.g. \cite{Gallager78}, \cite{Krichevsky},
\cite{Stubley}, \cite{Szpankowski00}, \cite{Reznik04}. Analysis of
their complexity can be found in \cite{Tjalkens00},
\cite{Tjalkens05}. Many useful practical implementation techniques
for such codes have also been reported, see, e.g.~\cite{Abrahams98},
\cite{MoffatTurpin},\cite{BrodnikCarlsson}.

When the source is not known, the best option available is to design
a {\em universal code\/} $\phi^*$ that minimize the worst case redundancy
for a class of sources~\cite{Fitingof66,Davisson73,Krichevsky}:
\begin{equation*}
R_{\phi^*}\left(n\right) = \inf_{\phi} \, \sup_{p}
\, R_{\phi}\left(n,p\right)\,.
\end{equation*}
An example of such a code can be constructed using the following estimates of words'
probabilities\footnote{This formula is due to Krichevsky and Trofimov~\cite{KrichevskyTrofimov},
and it assures uniform (in $p$) convergence to true probabilities with $n \rightarrow \infty$.
See~\cite{Krichevsky98} and \cite{XieBarron} for discussions on its background and optimality.}:
\begin{equation}
P_{KT}\left(w\right) = \frac{\Gamma \left( {k+1/2} \right) \, \Gamma
\left( {n-k+1/2} \right)}{\pi \Gamma\left( n+1 \right)} \,,
\label{eq:kt}
\end{equation}
where $\Gamma(x)$ is a $\Gamma$-function, $k$ is the weight of word $w$, and $n$ is its length.

Finally, we might be in a situation when exact value of parameter of the source is not known,
but we can access a sequence of symbols $u$ produced by this source in the past. We will call
such a sequence a {\em sample\/}, and will assume that it is $|u| = t$ bits long.
The task here is to design a {\em set of codes\/} (indexed by different values of this sample)
$\phi^*_u$, such that their resulting {\em worst case average redundancy\/} is minimal:
\begin{equation}
R_{\phi^*_{u}}\left(n,t\right) = \inf_{\left\{\phi_{u}\right\}}
\sup_{p} \sum_{|u| = t} \Pr(u)
R_{\phi_u}\left(n,p\right). \label{eq:sample_uni_r}
\end{equation}
Such codes are called {\em sample-based\/} or {\em adaptive} universal block
codes~\cite{Krichevsky68,Krichevsky75,KrichevskyTrofimov}.

In this paper we will study a particular implementation of adaptive block codes utilizing
the following estimates of probabilities of words $w$ given a sample $u$:
\begin{equation}
P_{KT}\left(w|u\right) = \frac{P_{KT}\left(u\,w\right)}{P_{KT}\left(u\right)}
= \frac{\Gamma \left( {k+s+1/2} \right) \, \Gamma
\left( {n+t-k-s+1/2} \right)}{\Gamma \left( {s+1/2} \right) \,
\Gamma \left( {t-s+1/2} \right)}
~\frac{\Gamma\left(t+1 \right)}{\Gamma\left(n+1 \right)}\,, \label{eq:cond_kt}
\end{equation}
where $s$ is the weight of a sample $u$, and $t$ is its length.

\section{Performance of Adaptive Block Codes}

The idea and original analysis of sample-based codes utilizing
estimator (\ref{eq:cond_kt}) belong to R.~E.~Krichevsky~\cite{Krichevsky75}.
In particular, he has shown (cf.
\cite[Theorem~1]{KrichevskyTrofimov}, \cite[Theorem~3.4.1]{Krichevsky}), that
the average redundancy rate of an adaptive block code is asymptotically
\begin{equation}
R_{\phi^*_{u}}\left(n,t\right) \sim \frac{1}{2\,n} \log \frac{n+t}{t} \,, \label{eq:1}
\end{equation}
where $n$ is a block size, and $t$ is the size of samples.

From (\ref{eq:1}) it is clear, that by using samples of length $t = O(n)$ it is
possible to lower redundancy rate of such codes to $O\left(\frac{1}{n}\right)$, which
matches the order of redundancy rate of block codes for known sources
\cite{CoverThomas,Krichevsky,Szpankowski00}.

However, in order to be able to understand full potential of such codes it is desirable
to know more exact expression for their redundancy, including terms affected by the
choice of actual code-construction algorithm (such as Huffman, Shannon, etc).
The theorem below\footnote{This is a simple generalization of our previous result
for adaptive Shannon codes~\cite{ReznikSzpankowski03}.} offers such a refinement.

\begin{theorem}[Reznik \& Szpankowsky 2003]
The average redundancy rate of an adaptive block code $\phi^*_{u}$ has the
following asymptotic behavior ($n,t \rightarrow \infty$):
\begin{eqnarray}
\lefteqn{R_{\phi^*_{u}}\left(n,t,p\right) = \sum_{|u| = t}
\Pr(u) R_{\phi^*_u}\left(n,p\right)} \nonumber \\
& = &
\frac{1}{n} \left\{{ \frac{1}{2} \log
\frac{t+n}{t} + \Delta_{\phi^*_{u}}(n,t,p) + \frac{1-4\,p\,q}{24\,p q}
\,\frac{n}{t\,(t+n)} - \frac{1-3\,p q}{24\,p^2 q^2}
\,\frac{(n+2\,t)\,n}{t^2\,(t+n)^2}}\right. \nonumber \\
& & ~~~~~~~ \left.{ + O\left(\frac{1}{t^3} +
\frac{1}{n^3}\right) } \right\}\,,~~~~~ \label{eq:thm1}
\end{eqnarray}
where $n$ is a block size, and $t$ is a sample size, $p,q=1-p$ are probabilities
of symbols of the input source, and where
\begin{equation}
\Delta_{\phi^*_{u}}\left(n,t,p\right) = \sum_{|u|=t} \sum_{|w|=n} \Pr(u) \Pr(w) \left[ {
\left| \phi^*_u(w) \right| + \log P_{KT}\left(w|u\right) }\right]
\end{equation}
is the average redundancy of code $\phi^*_{u}$ with respect to estimated
distribution (\ref{eq:cond_kt}).
\end{theorem}

The exact behavior of $\Delta_{\phi^*_{u}}\left(n,t,p\right)$ is algorithm-specific,
but for a large class of minimum-redundancy techniques, which includes conventional
Huffman and Shannon codes, we can say that this term is bounded in magnitude
\begin{equation*}
\left|\Delta(n,t,S)\right| \leqslant 1\,,
\end{equation*}
and that it exhibits oscillating behavior, which may or may not be convergent to
some constant depending on the value of the parameter $p$ (cf. \cite{Szpankowski00},
\cite{DrmotaHwangSzpankowski}, \cite{ReznikSzpankowski03}).

We also notice, that for short $t$ and $n$ the redundancy of such codes
becomes affected by the next following term:
\begin{equation*}
\frac{1-4\,p\,q}{24\,p q} \,\frac{n}{t\,(t+n)}
\end{equation*}
which is a function of the parameter of the source $p$. We plot leading factor of this
term in Figure~1, and conclude that for short blocks/samples performance of such codes
becomes sensitive to the asymmetry of the source.

\begin{figure}
\centerline{\resizebox{2.6in}{!}{\includegraphics{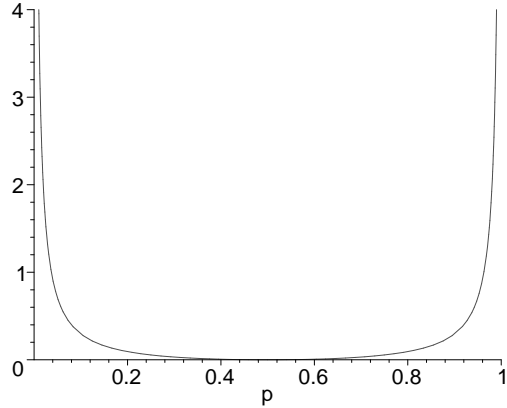}}}
\caption{Behavior of a factor $\frac{1-4\,p\,q}{24\,p q}$ in
redundancy expression (\ref{eq:thm1}).} \label{fig:fig1}
\end{figure}

Proof of this theorem can be found in Appendix A, and the rest of paper is devoted
to study of efficient algorithms for implementing such codes.

\section{Efficient Implementation of Block Codes}

We first notice, that in a memoryless model the probability of a word $w$
(or its estimate, cf. (\ref{eq:p_s}), (\ref{eq:kt}), (\ref{eq:cond_kt})) depends
only on its weight $k$, but not an actual pattern of its bits.
Hence, considering a set of all possible $n$-bit words, we can split it in $n+1$ groups:
\begin{equation}
\left\{0,1\right\}^n =  W_{n,0} \cup W_{n,1} \cup \ldots \cup W_{n,k} \cup \ldots \cup W_{n,n},
\end{equation}
containing words of the same weight $(k=0,\ldots,n)$, and the same probability.
As obvious, the sizes of such groups are $\left| W_{n,k} \right| = \binom{n}{k}$.
For further convenience, we will assume that each group $W_{n,k}$ stores words
in a lexicographic order. By $I_{n,k}(w)$ we will denote the index (position) of a word $w$
in a group $W_{n,k}$.

To describe the structure of our proposed mapping between words in groups $W_{n,k}$ and
their codewords, we will use an example code shown in Table 1.
This code was constructed using a modification of Huffman's
algorithm~\cite{Huffman}, in which additional steps were taken to ensure that
codewords located at the same level have same lexicographic order as input blocks
that they represent.
It is well-known that such a reordering is possible without any loss of compression
efficiency, and examples of prior algorithms that have been using this idea include
Huffman-Shannon-Fano codes~\cite{Connell}, canonic codes of Moffat
and Turpin~\cite{MoffatTurpin}, \cite{BrodnikCarlsson}, etc.

\begin{table}
\begin{center}
\caption{Example of a code constructed for 4-bit blocks with Bernoulli
 probabilities: \mbox{~~~~~~~~~~~~}$p^k q^{n-k}$, $p=0.9$.} \label{tab:1}
\vspace{0.05in}
{\small
\begin{tabular}
{|c|c|c|c|c|l|c|}\hline \hline
 {\centering\textbf{Block{$^{^{^{^{~}}}}$}$w$}} & {\centering\textbf{$k$}} &
{\centering\textbf{$I_{n,k}(w)$}} & {\centering\textbf{$\Pr(w)$}} &
{\centering\textbf{Length}} & {\centering\textbf{Code $\phi(w)$}} & {\centering\textbf{Sub-group}} \\
\hline
0000  &  0  & 0  & 0.6561  & 1  & 1          & 0 \\
0001  &  1  & 0  & 0.0729  & 3  & 001        & 1 \\
0010  &  1  & 1  & 0.0729  & 3  & 010        & 1 \\
0011  &  2  & 0  & 0.0081  & 6  & 000011     & 3 \\
0100  &  1  & 2  & 0.0729  & 3  & 011        & 1 \\
0101  &  2  & 1  & 0.0081  & 7  & 0000001    & 4 \\
0110  &  2  & 2  & 0.0081  & 7  & 0000010    & 4 \\
0111  &  3  & 0  & 0.0009  & 9  & 000000001  & 5 \\
1000  &  1  & 3  & 0.0729  & 4  & 0001       & 2 \\
1001  &  2  & 3  & 0.0081  & 7  & 0000011    & 4 \\
1010  &  2  & 4  & 0.0081  & 7  & 0000100    & 4 \\
1011  &  3  & 1  & 0.0009  & 9  & 000000010  & 5 \\
1100  &  2  & 5  & 0.0081  & 7  & 0000101    & 4 \\
1101  &  3  & 2  & 0.0009  & 9  & 000000011  & 5 \\
1110  &  3  & 3  & 0.0009  & 10 & 0000000001 & 6 \\
1111  &  4  & 0  & 0.0001  & 10 & 0000000000 & 7 \\
\hline \hline
\end{tabular}
}
\end{center}
\end{table}

\begin{figure}
\centerline{\resizebox{5in}{!}{\includegraphics{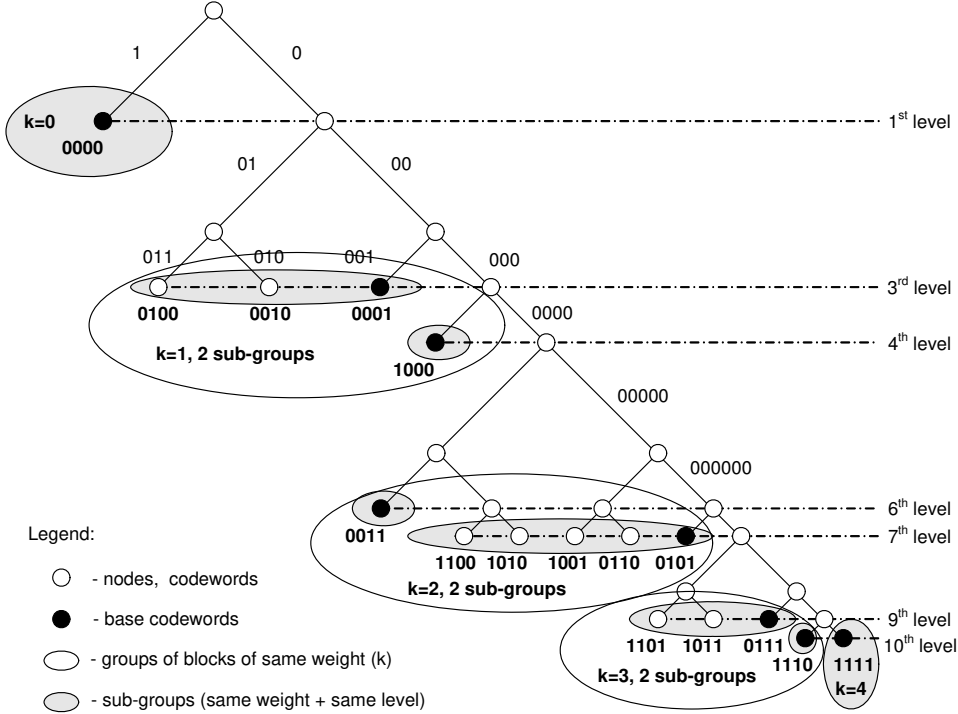}}}
\caption{Structure of an example block code.} \label{fig:fig1}
\end{figure}

In Figure 2 we depict the structure of this code. As expected, each group $W_{n,k}$
consists of at most two {\em sub-groups\/} containing codewords of the same length:
\footnote{This follows from the fact that all words in $W_{n,k}$ have the same probability,
and so-called {\em sibling property\/} of Huffman codes (cf.~\cite{Gallager78}, \cite{Stubley},
\cite{Szpankowski00}). This observation also holds true for 
Generalized Shannon codes~\cite{DrmotaSpa02LATIN} and possibly some other algorithms.}
\begin{equation}
W_{n,k} = W_{n,k,\ell} \cup W_{n,k,\ell+1}\,,
\end{equation}
where $\ell$ is the shortest code length that can be assigned to blocks
from $W_{n,k}$. Moreover, since words within $W_{n,k}$ group follow lexicographic order, then the
split between $W_{n,k,\ell}$ and $W_{n,k,\ell+1}$ is simply:
\begin{eqnarray}
W_{n,k,\ell} & = & \left\{ w \in W_{n,k}: I_{n,k}(w) < n_k \right\}\,, \\
W_{n,k,\ell+1} & = & \left\{ w \in W_{n,k}: I_{n,k}(w) \geqslant n_k \right\}\,,
\end{eqnarray}
where $n_k$ denotes the size of a subgroup with shorter codewords.

We will call lexicographically smallest codewords in each subgroup {\em base codewords\/}:
\begin{eqnarray}
B_{n,k,\ell} & = & \phi(w_0)\,, \\
B_{n,k,\ell+1} & = & \phi(w_{n_k})\,,
\end{eqnarray}
where $w_i:$ is $i$-th block in $W_{n,k}$, and note that the remaining codewords
in both subgroups can be computed as follows:
\begin{equation}
\phi(w_i) = \left[ \begin{array}{ll}
B_{n,k,\ell} + i, & \mbox{if $i < n_k $}\,, \\
B_{n,k,\ell+1} + i - n_k, & \mbox{if $i \geqslant n_k$}\,.
\end{array}\right. \label{eq:code1}
\end{equation}

We point out that such base codewords are only defined for non-empty sub-groups,
and that the number of such subgroups $S$ in a tree constructed for $n$-bit blocks is within:
\begin{equation}
n+1 \leqslant S \leqslant 2\,n\,.
\end{equation}

We also notice that multiple subgroups can reside on the same level
\footnote{This is one of the most obvious differences between our algorithm and
Connell~\cite{Connell}, or Moffat and Turpin~\cite{MoffatTurpin} algorithms,
which assign unique base codewords for each level, but then they need an
$O(n\,2^n)$-large reordering table to work with such codes. Here, the entire
structure is $O(n^2)$ bits large. Also, unlike \cite{Tjalkens00},
\cite{Tjalkens05} our algorithm does not assume any particular order of
probabilities based on weight $k$. This way we can implement codes for universal densities
(\ref{eq:kt}), and (\ref{eq:cond_kt}).} (see e.g. level 10
in tree in Figure 2), and the number of such collocated sub-groups cannot be greater than $n+1$.

\begin{Algorithm}[t]
{\small
\begin{verbatim}

/* encoder structure: */
typedef struct {
  unsigned short  nk[N+1];        /* # of elements in first (n,k) subgroup */
  unsigned char   sg[N+1][2];     /* (k,j) -> subgroup index mapping */
  unsigned char   len[S];         /* subgroup -> code length mapping */
  unsigned int    base[S];        /* subgroup -> base codeword mapping */
} ENC;

/* block encoder: */
unsigned block_enc (unsigned w, ENC *enc, BITSTREAM *bs)
{
  unsigned i, j, k, len, code;

  k = weight(w);                  /* split w into (k,index) */
  i = index(n,k,w);
  if (i >= enc->nk[k]) {          /* find subgroup containing w */
    i -= enc->nk[k];              /* adjust index */
    j = enc->sg[k][1];
  } else
    j = enc->sg[k][0];
  code = enc->base[j] + i;        /* generate code */
  len = enc->len[j];
  put_bits(code, len, bs);        /* write code to bitstream */

  return k;
}
\end{verbatim}}
\caption{Direct construction of block codes.}
\end{Algorithm}

\subsection{Proposed Algorithm for Block Encoding/Decoding}

Based on the discussion above we can now define a simple algorithm for direct computation
of block codes.

We assume that parameters $n_k$ ($0 \leqslant k\leqslant n$) are available,
and that for each non-empty sub-group we can obtain its level $\ell$ and its
base codeword $B_{n,k,\ell}$. Then the process of
encoding a block $w$ is essentially a set of the following steps:
\begin{itemize}
\item using $w$ obtain its weight $k$, and index $I_{n,k}(w)$
\item if $I_{n,k}(w) < n_k$ use first subgroup $W_{n,k,\ell}$ otherwise pick $W_{n,k,\ell+1}$
\item retrieve base codeword and compute the code according to (\ref{eq:code1}).
\end{itemize}

A complete C-language code of such a procedure is presented as Algorithm 1 above.

It can be seen that memory-wise this algorithm needs only $S$ base codewords ($O(n)$-bit
long\footnote{We note that additional memory reduction is possible by storing
incremental values of base codewords -- this is discussed in a companion paper \cite{Reznik07}.}),
$n+1$ values $n_k$ ($O(n)$-bit long), $S$ code lengths ($O(\log n)$-bit long), and $2\,(n+1)$
subgroup indices ($O(\log n)$-bit long). Given the fact that $S=O(n)$, the entire
structure needs $O(n^2)$ bits.

In a particular implementation shown in Algorithm 1, and assuming, e.g. that $n=20$ and $S=32$,
the size of this structure becomes $244$ bytes - far less than $2^{20}$ words needed
to present this code in a form of a direct table.

We note that for reasonably short blocks (e.g. $n \leqslant 12\ldots16$) computation
of their weights and indices (functions {\tt weight(.)} and {\tt index(.,.)}
in Algorithm 1), can be a matter of a single lookup, in which case, the entire encoding
algorithm needs at most $1$ comparison, $2$ additions, and $4$ lookups.

For larger blocks, one can use
the following well-known combinatorial formula (cf. \cite{Mudrov65}, \cite{Babkin71},
\cite{Schalkwijk72}, \cite{Cover73}, \cite{Tjalkens00}, \cite{Tjalkens05}):
\begin{equation}
I_{n,k}(w) = \sum_{j=1}^n w_{j} \binom{n-j}{\sum_{k=j}^n w_{k}}\,,
\label{eq:index}
\end{equation}
where $w_{j}$ represent individual bits of the word $w$, and it is
assumed that $\binom{n}{k} = 0$ for all $k > n$.
In order to implement it, one could either pre-compute all binomial
coefficients up to level $n$ in Pascal's triangle, or compute them
dynamically, using the following simple identities:
\begin{equation*}
\binom{n-k}{k-1}=\frac{k}{n}\binom{n}{k}\,, ~~\mbox{and}~~
\binom{n-k}{k}=\frac{n-k}{n}\binom{n}{k}\,.
\end{equation*}
The implementation based on pre-computed coefficients requires
$\frac{n\,(n+1)}{2} = O\left(n^2\right)$ words ($O(n^3)$ bits) of memory, and $O(n)$
additions. Dynamic computation of coefficients will require $O(n)$ additions,
multiplications and divisions, but the entire process needs only few registers.
Additional discussion on complexity of index computation can be
found in \cite{Tjalkens05}.

We now turn our attention to the design of a decoder. Here, we will
also need parameters $n_k$, base codewords, and their lengths. For
further convenience (as it was suggested by Moffat and
Turpin~\cite{MoffatTurpin}) we will use left-justified versions of
base values:
\begin{equation}
B^{lj}_{n,k,\ell} = B_{n,k,\ell} ~2^ {T-\ell}\,,
\end{equation}
where $T$ is the length of a machine word ($T > \max \ell$). We will store
such left-justified values in a lexicographically decreasing order.
Then, the decoding process can be described as follows:
\begin{itemize}
\item find first (top-most) subgroup with $B^{lj}_{n,k,\ell}$ being less than last $T$ bits in bitstream,
\item decode index of a block $I_{n,k}(w)$ (based on (\ref{eq:code1})), and
\item produce reconstructed block using its weight $k$ and index.
\end{itemize}

A complete C-language code of such a procedure is presented as Algorithm 2.

\begin{Algorithm}[t]
{\small
\begin{verbatim}

/* decoder structure: */
typedef struct {
  unsigned short  nk[N+1];         /* # of elements in first (n,k) subgroup */
  struct {unsigned char k:7,j:1;} kj[S]; /* subgroup -> (k,j) mapping */
  unsigned char   len[S];          /* subgroup -> code length mapping */
  unsigned int    lj_base[S];      /* subgroup -> left-justified codewords */
} DEC;

/* block decoder: */
unsigned block_dec (unsigned *w, DEC *dec, BITSTREAM *bs)
{
  unsigned i, j, k, len, val;

  val = bitstream_buffer(bs);
  for (j=0; dec->lj_base[j]>val; j++) ; /* find a subgroup */
  len = dec->len[j];
  scroll_bitstream(len, bs);      /* skip decoded bits  */
  i = (val - dec->lj_base[j]) >> (32-len);
  k = dec->kj[j].k;               /* get weight */
  j = dec->kj[j].j;               /* get sub-group index */
  if (j)                          /* reconstruct index */
    i += dec->nk[k];
  *w = word(n,k,i);               /* generate i-th word in (n,k) group */

  return k;
}
\end{verbatim}}
\caption{Decoding of a block codes.}
\end{Algorithm}

We note that (besides using left-justified base words) this algorithm
has almost identical data structure. The only new elements here are weights $k$
and subgroup level indicators $j$ ($j=0$ if subgroup contains shorter codewords,
and $j=1$ otherwise). Memory-wise it has very similar characteristics.

The main decoding process requires between $1$ and $S$ comparisons and lookups to find
a subgroup, $1$ or $2$ additions, $1$ shift, $1$ extra comparison, and $3$ extra
lookups.

As in Moffat-Turpin algorithm \cite{MoffatTurpin} the number of steps needed
for finding a subgroup can be further reduced by placing base codewords in a binary search
tree or using an extra lookup table, but in both cases we need to use extra memory to
accomplish this.

We note, that at the end of the decoding process we also need to convert word's
weight $k$ and index $I_{n,k}(w)$ into its actual value (function {\tt word()}
in Algorithm~2).
If blocks are reasonably short, this can be accomplished by a simple lookup.
Otherwise, we can synthesize the word by using the enumeration formulae~(\ref{eq:index}).
Complexity-wise this process is similar to index computation in
the encoder.

\section{Design of an Adaptive Block Coder}
Using above described algorithms we can now define a system for adaptive
encoding/decoding of blocks of data.

In this system, we assume that input blocks can be encoded under the following conditions:
\begin{enumerate}
\item there is no context - i.e. we implement universal code,
\item the context is given by one previously seen block - i.e. $t=n$,
\item the context is given by two previously seen blocks - i.e. $t=2\,n$.
\end{enumerate}

We note, that instead of using actual blocks as contexts it is sufficient (due
to memoryless nature of the source) to use their weights.

This means, that for $t$-bit samples, we will need to have an array of $t+1$ code
structures indexed by their weights $s$. To further save space, we can use
symmetry of KT-distributions (\ref{eq:cond_kt}) with respect to $s$ and $k$:
replace $s=t-s$ and flip bits (i.e. force $k=n-k$) every time when $s>t/2$.
This way we will only need to define $t/2 + 1$ tables.

Hence, the overall amount of memory needed by our adaptive code becomes $1 + n/2 +1 + n+1 = 1.5\,n + 3$
tables. Specific memory estimates for block sizes $n=8 \ldots 20$, are shown in Table 2.

\begin{table}
\begin{center}
\caption{Memory usage estimates [in bytes] for different block sizes} \label{tab:2}
\vspace{0.05in}
{\small
\begin{tabular}
{|c|c|c|c|c|}\hline \hline
 {\centering\textbf{$n$}} & {\centering\textbf{$\max t$}} & {\centering\textbf{$\max S^{^{^{^{~}}}}$}} &
{\centering\textbf{Size of a single table}} & {\centering\textbf{Tables for all contexts}} \\
\hline
8   &  16  & 14  & 102  & 1530 \\
12  &  24  & 19  & 140  & 2940 \\
16  &  32  & 25  & 184  & 4968 \\
20  &  40  & 29  & 216  & 7128 \\
\hline \hline
\end{tabular}
}
\end{center}
\end{table}

In out test implementation we've generated all these tables using KT-estimated
densities (\ref{eq:kt}) and (\ref{eq:cond_kt}), and using modified Huffman code-
construction algorithm, as described in Section~3.

In Appendix B we provide a complete code of a program implementing such a system.

\section{Experimental Study of Performance of our Algorithm}

In this section we provide experimental results of evaluation of performance
of our adaptive code with block size $n=16$, and compare it with the following
well known algorithms:
\begin{itemize}
\item IBM's Q-coder algorithm \cite{QCoder} adopted in JBIG standard for
image compression~\cite{T82} (we've used implementation from JBIG's jbigkit);
\item CABAC binary arithmetic encoder \cite{CABAC} from MPEG AVC/ITU-T H.264
standard for video coding~\cite{H264}.
\end{itemize}

In order to conduct our tests we've used computed-generated sequences of bits
simulating output from a binary Bernoulli source with probability $p$.
Lengths of such sequences ranged from $16$ to $1024$, and for each particular
length we have generated $Q=1000000$ samples of such sequences.

Relative redundancy rates were computed as:
\begin{equation*}
\mathrm{Rate} = \frac{(\mathrm{sum~of~lengths~of~all~codes~produced~for~Q~sequences})/Q -H(p)}{H(p)}
\end{equation*}

For our adaptive code we've used the following structure of contexts:
\begin{itemize}
\item first $16$-bit block is encoded without context (universal code),
\item second block is encoded using first one as its context (code with $t=16$),
\item third and all subsequent blocks are encoded using two previous
blocks in a sequence as contexts (sample-based code with $t=32$).
\end{itemize}

\begin{figure}[t]
\centerline{\resizebox{2.6in}{!}{\includegraphics{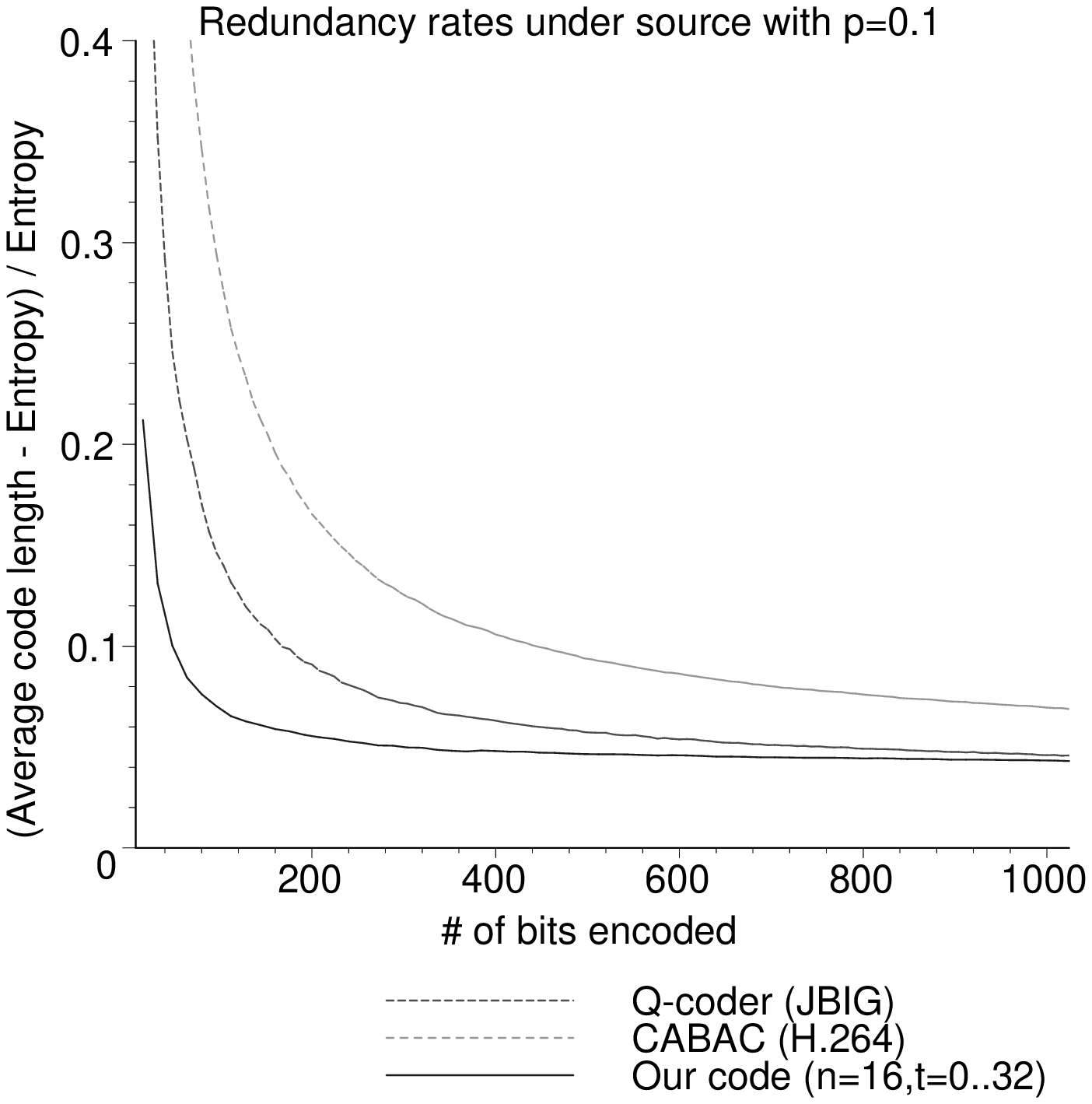}} ~~ \resizebox{2.6in}{!}{\includegraphics{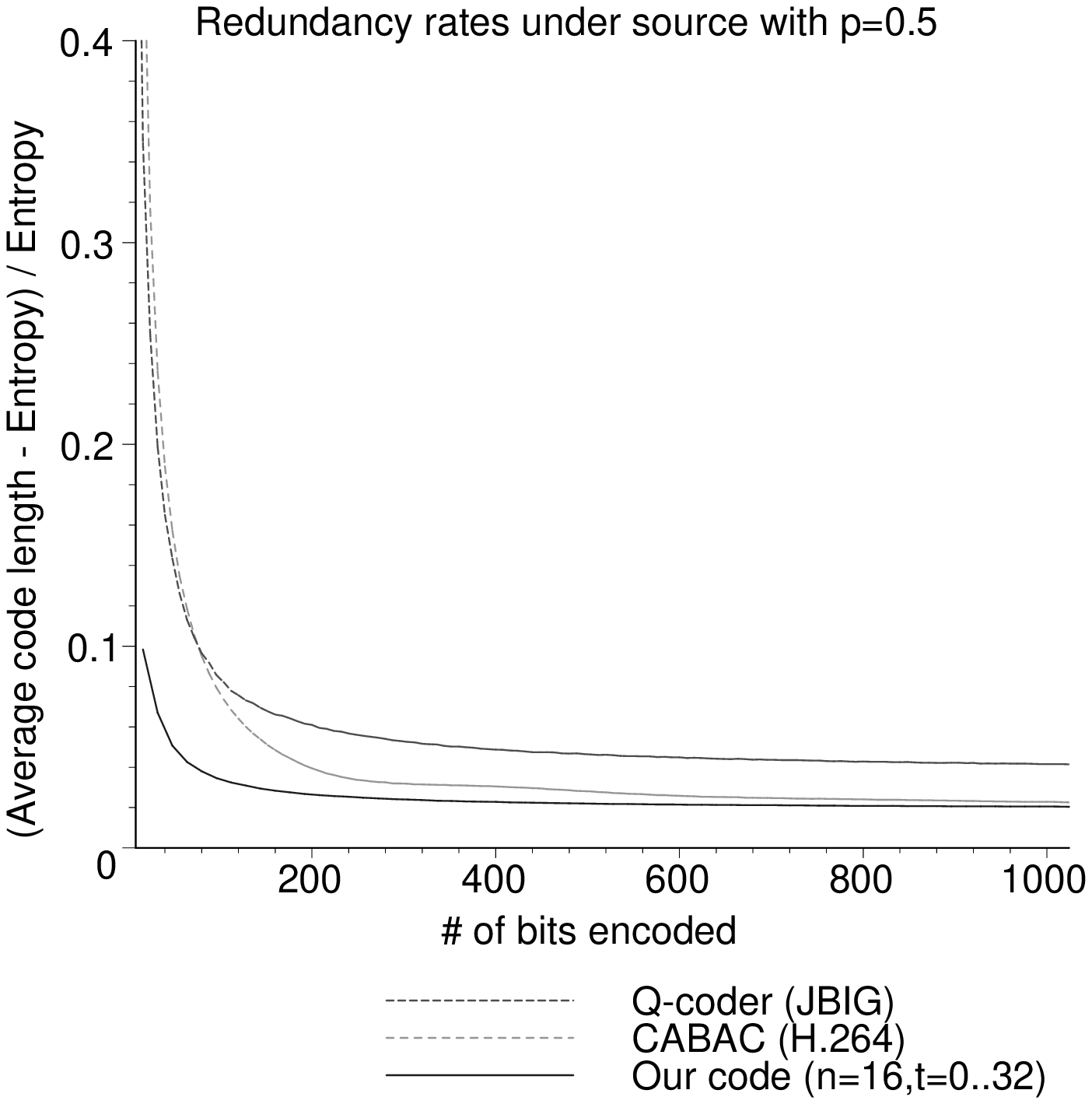}}}
\caption{Comparison of redundancy rates under memoryless sources with $p=0.1$ (left) and $p=0.5$ (unbiased case, right).} \label{fig:fig2}
\end{figure}

\begin{figure}
\centerline{\resizebox{3in}{!}{\includegraphics{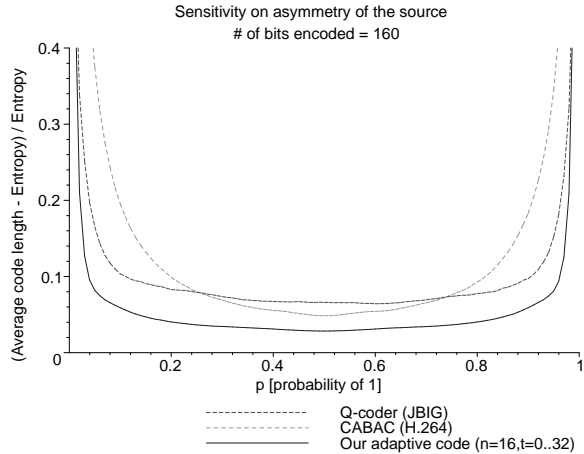}}}
\caption{Sensitivity of redundancy to asymmetry of the source.} \label{fig:fig3}
\end{figure}

The results of our experimental study are shown in Figures 3 and 4.
It can be seen that our code has a much faster rate of convergence than that of Q-coder
or CABAC algorithms. It clearly outperforms them for short sequences, and becomes
comparable to the best of these two when the total length of encoded bits
approaches~$1024$.

In Figure 4 we also show analysis of sensitivity of redundancy rates of these codes
to asymmetry of the source. Here, after $160$ encoded bits (or $10$ $16$-bit blocks)
our algorithm delivers much lower redundancy compared to others. Its behavior
is consistent with one that was predicted by our Theorem 1.

\newpage
\appendix
\section{Proof of Theorem 1}

We need to evaluate average redundancy of an adaptive code $\phi^*_{u}$
working with samples of length $t$ and blocks of length $n$ produced by a
binary memoryless source with parameter $p$:
\begin{equation}
R_{\phi^*_{u}}\left(n,t,p\right) = \frac{1}{n} \sum_{|u|=t} \sum_{|w|=n} \Pr(u) \Pr(w)
\left| \phi_u(w) \right| - H(p)\,. \label{eq:R1}
\end{equation}

We will further assume that each codeword $\phi_u(w)$ is generated on the basis
of KT-estimated probability $P_{KT}\left(w|u\right)$ (\ref{eq:cond_kt}), and therefore
we can rewrite (\ref{eq:R1}) as follows:
\begin{equation}
R_{\phi^*_{u}}\left(n,t,p\right) = \frac{1}{n} \sum_{|u|=t} \sum_{|w|=n} \Pr(u) \Pr(w)
\log P^{-1}_{KT}\left(w|u\right) - H(p) + \frac{1}{n} \Delta_{\phi^*_{u}}\left(n,t,p\right)\,,~ \label{eq:R2}
\end{equation}
where by
\begin{equation}
\Delta_{\phi^*_{u}}\left(n,t,p\right) = \sum_{|u|=t} \sum_{|w|=n} \Pr(u) \Pr(w) \left[ {
\left| \phi_u(w) \right| + \log P_{KT}\left(w|u\right) }\right]
\end{equation}
we denote the redundancy of code $\phi^*_{u}$ with respect to the distribution it implements.

We know, that given density $P_{KT}\left(w|u\right)$ most existing minimum redundancy
block codes (such as block Huffman, or Shannon algorithms) produce codewords such that:
\begin{equation*}
\left \lfloor \log P_{KT}\left(w|u\right) \right\rfloor
 \leqslant \left| \phi_u(w) \right| \leqslant \left \lceil \log P_{KT}\left(w|u\right) \right\rceil\,,
\end{equation*}
which implies that $\Delta_{\phi^*_{u}}\left(n,t,p\right)$ is a quantity of bounded magnitude:
\begin{equation*}
\left|{\Delta_{\phi^*_{u}}\left(n,t,p\right)}\right| \leqslant 1\,,
\end{equation*}
and which might have some erratic or oscillating behavior (cf. \cite{Szpankowski00}, \cite{DrmotaHwangSzpankowski}).

We now concentrate our attention on the main sum in (\ref{eq:R2}):
\begin{eqnarray}
\lefteqn{-\sum_{|u|=t} \sum_{|w|=n} \Pr(u) \Pr(w) \log
P_{KT}\left(w|u\right) = } \nonumber \\
& & = - \sum_{|u\,w|=t+n} \Pr(u\,w) \log P_{KT}\left(u\,w\right) +
\sum_{|u|=t}\Pr(u)\log P_{KT}\left( u \right) \nonumber \\
& & = (t+n)\,C_{KT}(t+n,p) - t\, C_{KT}(t,p) \,, \label{eq:C3}
\end{eqnarray}
where
\begin{equation} C_{KT}(n,p) = - \frac{1}{n} \sum_{|w|=n}\Pr(w)\log P_{KT}\left( w
\right)\,, \label{eq:C_KT}
\end{equation}
is the average rate of the KT-estimator processing $n$-symbols words produced
by $p$.

\subsection{Asymptotic average rates of empirical entropy and KT-estimator}
Consider KT-estimated probability of a word $w$
\begin{equation*}
P_{KT}\left(w\right) = \frac{\Gamma \left( {k+1/2} \right) \, \Gamma \left(
{n-k+1/2} \right)}{\pi \Gamma\left( n+1 \right)} \,. \label{eq:kt_nk}
\end{equation*}

Using Stirling's approximation (and excluding cases when $k = 0,n$), we can
show that:
\begin{eqnarray}
-\log P_{KT}\left(w\right) & = & n\, F\left( w \right) + \frac{1}{2} \log n +
\frac{1}{2} \log \frac{\pi}{2} + \frac{1}{12\,n} \nonumber \\
& & + \frac{1}{24\,k} +\frac{1}{24\,(n-k)} + O \left(\frac{1}{k^3} +
\frac{1}{(n-k)^3}\right) \,, \label{eq:log_kt}
\end{eqnarray}
where:
\begin{equation}
F\left(w\right) = -\frac{k}{n} \log \left(\frac{k}{n}\right) -\frac{n-k}{n}
\log \left(\frac{n-k}{n}\right) \,.
\end{equation}
is an empirical entropy \cite{Fitingof66,KrichevskyTrofimov} of a word $w$.

The average rate of the empirical entropy $F(w)$ under source $p$ is:
\begin{eqnarray}
\lefteqn{\sum_{|w|=n}\Pr(w)F(w)} \nonumber \\
& = & -\sum_{k=0}^n \binom{n}{k} p^k q^{n-k} \left[ { \frac{k}{n} \log
\left(\frac{k}{n}\right) +\frac{n-k}{n} \log \left(\frac{n-k}{n}\right)
}\right] \nonumber \\
& = & \log n - \sum_{k=1}^n \binom{n}{k} p^{k} q^{n-k}  \frac{k}{n} \log(k) -
\sum_{k=0}^{n-1} \binom{n}{k} p^{k} q^{n-k} \frac{n-k}{n} \log
\left(n-k\right) \nonumber \\
& = &  \log n - p\,\sum_{k=1}^n \binom{n-1}{k-1} p^{k-1} q^{n-k} \log(k) -
q\,\sum_{k=0}^{n-1} \binom{n-1}{k} p^k q^{n-1-k} \log \left(n-k \right)
\nonumber\\
& = & \log n - p\,f(n-1,k,p) - q\,f(n-1,k,q)\,. \label{eq:e_F}
\end{eqnarray}
where:
\begin{equation}
f(n,k,\theta) =  \sum_{k=0}^{n} {n \choose k} \theta^{k} (1-\theta)^{n-k}
\log\left(1+k\right)\,. \label{eq:f}
\end{equation}

We immediately notice that (\ref{eq:f}) belongs to a class of so-called binomial
sums (see, e.g.~\cite[p.~92]{Borovkov}), and that for large $n$ we must have
(a uniform in $\theta$) convergence $f(n,k,\theta)
\rightarrow \log \left(1+\theta\,n\right)$. However, in order to obtain a more detailed
asymptotic of $f(n,k,\theta)$, one must use analytic techniques, such
as analytic depoissonization~\cite{JacquetSzpankowski96,
JacquetSzpankowski98}, or singularity analysis of generating
functions~\cite{FlajoletOdlyzko}.

In fact, the last approach in application
to a class of polylogarithmic Bernoulli sums has already been used
by P. Flajolet \cite{Flajolet99}, and in particular, he has shown that
\begin{equation*}
\sum_{k=1}^n \binom{n}{k} \theta^{k} (1-\theta)^{n-k} \log k = \log \left(
\theta\,n \right) + \frac{\theta - 1}{2\,\theta\,n} - \frac{\theta^2 -
6\,\theta + 5}{12\,\theta^2n^2} + O\left(\frac{1}{n^3}\right)\,,
\label{eq:flajolet}
\end{equation*}
which is a very similar sum to one that we need to evaluate (\ref{eq:f}). To
take advantage of this existing result, we simply replace $\log(1+k)$ in
(\ref{eq:f}) with:
\begin{equation}
\log(k+1) = \log(k) + \frac{1}{k+1} + \frac{1}{2} \frac{1}{(k+1)(k+2)} +
\frac{5}{6} \frac{1}{(k+1)(k+2)(k+3)} + \ldots \,. \label{eq:log_fact_exp}
\end{equation}
The evaluation of the sums containing factorial powers of $k$ yields:
\begin{eqnarray}
S_1(n,\theta) & = & \sum_{k=1}^{n} \binom{n}{k} \theta^{k} (1-\theta)^{n-k}
\frac{1}{k+1} \nonumber \\
& = & \frac{1}{\theta\,(n+1)}\left[{1 - (1-\theta)^n -\theta\,n\,(1-\theta)^n
}\right] \nonumber \\
& = & \frac{1}{\theta\,n} - \frac{1}{\theta\,n^2} +
O\left(\frac{1}{n^3}\right)
\,, \label{eq:S_1} \\
S_2(n,\theta) & = & \sum_{k=1}^{n} \binom{n}{k} \theta^{k} (1-\theta)^{n-k}
\frac{1}{(k+1)(k+2)} \nonumber \\
& = & \frac{1}{\theta^2\,(n+1)\,(n+2)}\left[{1 - (1-\theta)^n -
\theta\,n\,(1-\theta)^n - \frac{\theta^2\,n\,(n+1)}{2}(1-\theta)^n}\right] \nonumber \\
& = &  \frac{1}{\theta^2\,n^2} + O\left(\frac{1}{n^3} \right)
\,, \label{eq:S_2}
\end{eqnarray}
and it is clear that the contribution of the subsequent terms
in~(\ref{eq:log_fact_exp}) to the sum~(\ref{eq:f}) is within
$O\left(\frac{1}{n^3} \right)$.

Combining the above formulae, we obtain
\begin{equation}
f(n,k,\theta) = \log \left( \theta\,n \right) + \frac{1 + \theta}
{2\,\theta\,n} - \frac{\theta^2 + 6\,\theta - 1}{12\,\theta^2n^2} +
O\left(\frac{1}{n^3}\right)\,, \label{eq:f_final}
\end{equation}
and subsequently (after plugging (\ref{eq:f_final}) in (\ref{eq:e_F}) and
some simple algebra):
\begin{equation}
\sum_{|w|=n}\Pr(w)F(w) = H(p) -\frac{1}{2\,n} + \frac{pq-1}{12\,p\,q\,n^2} +
O\left(\frac{1}{n^3}\right) \,, \label{eq:e_F_final}
\end{equation}
which is an (up to $O\left(\frac{1}{n^3}\right)$-term) accurate asymptotic
expression for the average rate of empirical entropy.

\begin{remark}
The average rate of empirical entropy has already been studied by Krichevsky
(cf. \cite[Lemma~1]{KrichevskyTrofimov} and \cite[Lemma~3.2.1]{Krichevsky}).
His conclusion was that:
\begin{equation*}
-\frac{1}{n} \leqslant \sum_{|w|=n}\Pr(w)F(w) - H(p) \leqslant 0 \,.
\end{equation*}
Our formula (\ref{eq:e_F_final}) confirms and refines this statement.
\end{remark}

We now focus our attention on the average rate of the
KT-estimator~(\ref{eq:C_KT}). Using our asymptotic expression
(\ref{eq:log_kt}) and replacing $\frac{1}{k}$ and $\frac{1}{n-k}$ with the
appropriate factorial powers
we can show that:
\begin{eqnarray}
C_{KT}(n,p) & = & - \frac{1}{n} \sum_{|w|=n}\Pr(w)\log P_{KT}\left( w \right)
\nonumber \\
& = & \sum_{|w|=n}\Pr(w) F(w) + \frac{1}{n} \left\{ {\frac{1}{2}\log n +
\frac{1}{2} \log \frac{\pi}{2} + \frac{1}{12\,n} } \right. \nonumber \\
& &~~~~~~~~~~ \left.{ + \frac{1}{24} \left[ {S_1(n,p) + S_2(n,p)} \right] +
\frac{1}{24} \left[ {S_1(n,q) + S_2(n,q)} \right]} \right\} \nonumber \\
& &~~~~~~~~~~ + O \left(\frac{1}{n^4} \right) \,, \label{eq:C_KT_2}
\end{eqnarray}
where $S_1(n,\theta)$ and $S_2(n,\theta)$ are already familiar sums
(\ref{eq:S_1}) and (\ref{eq:S_2}).

Now by using (\ref{eq:e_F_final}) and expanding all the expressions in
(\ref{eq:C_KT_2}) we finally obtain:
\begin{eqnarray}
C_{KT}(n,p) & = & H(p) + \frac{1}{2\,n} \left\{ {\log n + \log \frac{\pi}{2}
-1 - \frac{1-4\,p\,q}{12\,p q\,n} + \frac{1-3\,p q}{12\,p^2 q^2\,n^2} }
\right\} \nonumber \\
& & + O \left(\frac{1}{n^4} \right) \,. \label{eq:C_KT_final}
\end{eqnarray}

\subsection{Asymptotic average rate of the adaptive block code}
Using (\ref{eq:R2}-\ref{eq:C_KT}) we can now say that
\begin{equation}
R_{\phi^*_{u}}\left(n,t,p\right) = \frac{1}{n} \left[ {(t+n)\,C_{KT}(t+n,p) - t\,
C_{KT}(t,p) - n H(p) + \Delta_{\phi^*_{u}}(n,t,p)} \right],
\end{equation}
where $C_{KT}(n,p)$ is the average rate of the KT-estimator (\ref{eq:C_KT}).

By applying our asymptotic result (\ref{eq:C_KT_final}) for $C_{KT}(n,p)$ and
combining the remaining (after some cancellations) terms we arrive at
\begin{eqnarray*}
R_{\phi^*_{u}}\left(n,t,p\right) & = & \frac{1}{n} \left\{{ \frac{1}{2} \log
\frac{t+n}{t} + \Delta_{\phi^*_{u}}(n,t,p) + \frac{1-4\,p\,q}{24\,p q}
\,\frac{n}{t\,(t+n)}  }\right. \nonumber \\
& & ~~~~~~~~~~~~~~ \left.{ - \frac{1-3\,p q}{24\,p^2 q^2}
\,\frac{(n+2\,t)\,n}{t^2\,(t+n)^2} + O\left(\frac{1}{t^3} +
\frac{1}{n^3}\right) } \right\} \,,
\end{eqnarray*}
which is formula (\ref{eq:thm1}) claimed by our theorem.

\section{Example implementation of adaptive block coder}



{\tiny
\begin{verbatim}
/* bitstream.h: */

typedef struct _BITSTREAM BITSTREAM;

void bitstream_open(BITSTREAM *p, unsigned char *pbs, unsigned bit_offset, int read);
void bitstream_close(BITSTREAM *p, unsigned char **p_pbs, unsigned *p_bit_offset, int write);

void put_bits(unsigned bits, int len, BITSTREAM *p);
unsigned bitstream_buffer(BISTREAM *p);
void scroll_bitstream(int len, BITSTREAM *p);


/* blade.h: */

/* encoder functions: */
void blade_enc_init(void);
unsigned blade_enc_0(unsigned block, BITSTREAM *bs);
unsigned blade_enc_1(unsigned block, unsigned cx, BITSTREAM *bs);
unsigned blade_enc_2(unsigned block, unsigned cx1, unsigned cx2, BITSTREAM *bs);

/* decoder functions: */
void blade_dec_init(void);
unsigned blade_dec_0(unsigned *block, BITSTREAM *bs);
unsigned blade_dec_1(unsigned *block, unsigned cx, BITSTREAM *bs);
unsigned blade_dec_2(unsigned *block, unsigned cx1, unsigned cx2, BITSTREAM *bs);


/* blade_12.c:  implements 12-bit BLADE encoder/decoder */

#define N   12      /* block size */
#define SGS 19      /* max # of subgroups */

/* encoder structure: */
typedef struct {
  unsigned short  nk [N+1];        /* # of elements in first (n,k) subgroup */
  unsigned char   len [SGS];       /* subgroup -> code length mapping */
  unsigned char   sg [N+1][2];     /* (k,j) -> subgroup index mapping */
  unsigned int    base [SGS];      /* subgroup -> base codeword mapping */
} BLADE_ENC;

/* w -> (k,index) mapping: */
static struct {unsigned short k:4, i:12;} w_ki[1<<N];

/*
 * BLADE encoder:
 * Returns:
 *  # of bits set in encoded pattern
 */
unsigned blade_enc (unsigned w, BLADE_ENC *enc, BITSTREAM *bs)
{
  unsigned i, j, k, len, code;

  k = w_ki[w].k;                  /* split w into (k,index) */
  i = w_ki[w].i;
  if (i >= enc->nk[k]) {          /* find subgroup containing w */
    i -= enc->nk[k];              /* adjust index */
    j = enc->sg[k][1];
  } else
    j = enc->sg[k][0];
  code = enc->base[j] + i;        /* generate code */
  len = enc->len[j];
  put_bits(code, len, bs);

  return k;
}

/* decoder structure: */
typedef struct {
  unsigned int    sgs;             /* number of subgroups */
  unsigned short  nk [N+1];        /* # of elements in first (n,k) subgroup */
  unsigned char   len [SGS];       /* subgroup -> code length mapping */
  struct {unsigned char k:7,j:1;} kj [SGS]; /* subgroup -> (k,j) mapping */
  unsigned int    lj_base [SGS];   /* subgroup -> left-justified codewords */
} BLADE_DEC;

/* (k,index) -> w mapping:*/
static unsigned short *ki_w[N+1], _w[1<<N];

/*
 * BLADE decoder:
 * Returns:
 *  # of bits set in encoded pattern
 */
unsigned blade_dec (unsigned *w, BLADE_DEC *dec, BITSTREAM *bs)
{
  unsigned i, j, k, len, val;

  val = bitstream_buffer(bs);
  for (j=0; j<dec->sgs; j++)      /* find subgroup */
    if (dec->lj_base[j] <= val)
      break;
  len = dec->len[j];
  scroll_bitstream(len, bs);      /* skip decoded bits  */
  i = (val - dec->lj_base[j]) >> (32-len);
  k = dec->kj[j].k;
  j = dec->kj[j].j;
  if (j)                          /* convert to (n,k)-group's index */
    i += dec->nk[k];
  *w = ki_w[k][i];                /* produce reconstructed block */

  return k;
}

/**********************************************
* Pre-computed BLADE decoder tables:
*/
static BLADE_DEC dec_t [1+(N/2+1)+(N+1)] = {
  { /* no context/ universal code: */ 15,
  {1,12,66,92,495,792,924,792,495,122,66,12,1}, {3,3,7,7,10,10,11,11,12,12,13,13,14,14,14},
  {{0,0},{12,0},{1,0},{11,0},{2,0},{10,0},{3,0},{9,0},{3,1},{9,1},{4,0},{8,0},{5,0},{6,0},{7,0}},
  {0xE0000000,0xC0000000,0xA8000000,0x90000000,0x7F800000,0x6F000000,0x63800000,0x54400000,
  0x4C400000,0x46200000,0x36A80000,0x27300000,0x1AD00000,0x0C600000,0x00000000} },
  { /* (12,0): */ 17,
  {1,8,66,64,495,792,924,792,334,220,66,11,1}, {1,5,6,9,12,13,15,17,19,20,21,22,22,23,23,24,24},
  {{0,0},{1,0},{1,1},{2,0},{3,0},{3,1},{4,0},{5,0},{6,0},{7,0},{8,0},{8,1},{9,0},{10,0},{11,0},{11,1},{12,0}},
  {0x80000000,0x40000000,0x30000000,0x0F000000,0x0B000000,0x06200000,0x02420000,0x00B60000,
  0x00428000,0x00110000,0x00069000,0x00040C00,0x00009C00,0x00001800,0x00000200,0x00000100,0x00000000} },
  { /* (12,1): */ 16,
  {1,12,17,220,495,792,924,340,495,220,66,10,1}, {2,5,8,9,11,13,15,16,17,18,18,19,19,19,19,20},
  {{0,0},{1,0},{2,0},{2,1},{3,0},{4,0},{5,0},{6,0},{7,0},{7,1},{8,0},{9,0},{10,0},{11,0},{12,0},{11,1}},
  {0xC0000000,0x60000000,0x4F000000,0x36800000,0x1B000000,0x0B880000,0x05580000,0x01BC0000,
  0x01120000,0x00A10000,0x00254000,0x0009C000,0x00018000,0x00004000,0x00002000,0x00000000} },
  { /* (12,2): */ 15,
  {1,12,66,211,495,792,924,792,486,220,66,12,1}, {3,6,8,10,11,12,14,15,16,16,17,17,17,17,17},
  {{0,0},{1,0},{2,0},{3,0},{3,1},{4,0},{5,0},{6,0},{7,0},{8,0},{8,1},{9,0},{10,0},{11,0},{12,0}},
  {0xE0000000,0xB0000000,0x6E000000,0x39400000,0x38200000,0x19300000,0x0CD00000,0x05980000,
  0x02800000,0x009A0000,0x00958000,0x00278000,0x00068000,0x00008000,0x00000000} },
  { /* (12,3): */ 16,
  {1,12,30,220,495,792,924,792,19,220,6,12,1}, {4,6,8,9,10,12,13,14,14,14,14,14,14,15,15,15},
  {{0,0},{1,0},{2,0},{2,1},{3,0},{4,0},{5,0},{6,0},{7,0},{8,0},{10,0},{11,0},{12,0},{8,1},{10,1},{9,0}},
  {0xF0000000,0xC0000000,0xA2000000,0x90000000,0x59000000,0x3A100000,0x21500000,0x12E00000,
  0x06800000,0x06340000,0x061C0000,0x05EC0000,0x05E80000,0x02300000,0x01B80000,0x00000000} },
  { /* (12,4): */ 16,
  {1,12,66,220,495,303,924,792,495,219,66,4,1}, {5,7,9,10,12,12,12,12,13,13,13,13,13,13,14,14},
  {{0,0},{1,0},{2,0},{3,0},{4,0},{5,0},{11,0},{12,0},{5,1},{11,1},{6,0},{7,0},{9,0},{10,0},{9,1},{8,0}},
  {0xF8000000,0xE0000000,0xBF000000,0x88000000,0x69100000,0x56200000,0x55E00000,0x55D00000,
  0x46880000,0x46480000,0x29680000,0x10A80000,0x09D00000,0x07C00000,0x07BC0000,0x00000000} },
  { /* (12,5): */ 15,
  {1,12,66,220,495,792,509,792,350,220,66,12,1}, {6,8,10,10,11,11,12,12,12,12,12,12,13,13,13},
  {{0,0},{1,0},{2,0},{12,0},{3,0},{11,0},{4,0},{5,0},{6,0},{8,0},{9,0},{10,0},{6,1},{8,1},{7,0}},
  {0xFC000000,0xF0000000,0xDF800000,0xDF400000,0xC3C00000,0xC2400000,0xA3500000,0x71D00000,
  0x52000000,0x3C200000,0x2E600000,0x2A400000,0x1D480000,0x18C00000,0x00000000} },
  { /* (12,6): */ 15,
  {1,12,66,47,495,792,924,792,495,85,66,12,1}, {8,8,9,9,11,11,11,11,12,12,12,12,12,12,13},
  {{0,0},{12,0},{1,0},{11,0},{2,0},{3,0},{9,0},{10,0},{3,1},{9,1},{4,0},{5,0},{7,0},{8,0},{6,0}},
  {0xFF000000,0xFE000000,0xF8000000,0xF2000000,0xE9C00000,0xE3E00000,0xD9400000,0xD1000000,
  0xC6300000,0xBDC00000,0x9ED00000,0x6D500000,0x3BD00000,0x1CE00000,0x00000000} },
  { /* (24,0): */ 19,
  {1,12,25,220,487,791,924,787,494,220,66,11,1}, {1,5,9,10,13,16,17,19,20,22,24,25,26,27,28,30,31,32,32},
  {{0,0},{1,0},{2,0},{2,1},{3,0},{4,0},{4,1},{5,0},{5,1},{6,0},{7,0},{7,1},{8,0},{8,1},{9,0},{10,0},{11,0},{11,1},{12,0}},
  {0x80000000,0x20000000,0x13800000,0x09400000,0x02600000,0x00790000,0x00750000,0x00122000,0x00121000,
  0x0003A000,0x00008D00,0x00008A80,0x00000F00,0x00000EE0,0x00000120,0x00000018,0x00000002,0x00000001,0x00000000} },
  { /* (24,1): */ 17,
  {1,7,66,220,495,326,924,792,495,4,66,11,1}, {1,5,6,9,12,15,17,18,20,22,23,24,25,26,27,28,28},
  {{0,0},{1,0},{1,1},{2,0},{3,0},{4,0},{5,0},{5,1},{6,0},{7,0},{8,0},{9,0},{9,1},{10,0},{11,0},{11,1},{12,0}},
  {0x80000000,0x48000000,0x34000000,0x13000000,0x05400000,0x01620000,0x00BF0000,0x004A8000,
  0x0010C000,0x00046000,0x00008200,0x00007E00,0x00001200,0x00000180,0x00000020,0x00000010,0x00000000} },
  { /* (24,2): */ 17,
  {1,12,47,220,495,792,924,1,495,220,58,11,1}, {2,5,8,9,11,14,16,18,19,20,21,22,23,24,24,25,25},
  {{0,0},{1,0},{2,0},{2,1},{3,0},{4,0},{5,0},{6,0},{7,0},{7,1},{8,0},{9,0},{10,0},{10,1},{11,0},{11,1},{12,0}},
  {0xC0000000,0x60000000,0x31000000,0x27800000,0x0C000000,0x04440000,0x012C0000,0x00450000,
  0x0044E000,0x00137000,0x0003F800,0x00008800,0x00001400,0x00000C00,0x00000100,0x00000080,0x00000000} },
  { /* (24,3): */ 17,
  {1,6,66,1,495,4,924,792,495,220,66,7,1}, {2,5,6,8,10,11,13,14,15,16,18,19,20,21,21,22,22},
  {{0,0},{1,0},{1,1},{2,0},{3,0},{3,1},{4,0},{5,0},{5,1},{6,0},{7,0},{8,0},{9,0},{10,0},{11,0},{11,1},{12,0}},
  {0xC0000000,0x90000000,0x78000000,0x36000000,0x35C00000,0x1A600000,0x0AE80000,0x0AD80000,
  0x04B00000,0x01140000,0x004E0000,0x00102000,0x00026000,0x00005000,0x00001800,0x00000400,0x00000000} },
  { /* (24,4): */ 15,
  {1,12,66,220,495,10,924,792,495,220,66,7,1}, {3,6,8,10,12,13,14,15,16,17,18,19,19,20,20},
  {{0,0},{1,0},{2,0},{3,0},{4,0},{5,0},{5,1},{6,0},{7,0},{8,0},{9,0},{10,0},{11,0},{11,1},{12,0}},
  {0xE0000000,0xB0000000,0x6E000000,0x37000000,0x18100000,0x17C00000,0x0B880000,0x04500000,
  0x01380000,0x00408000,0x00098000,0x00014000,0x00006000,0x00001000,0x00000000} },
  { /* (24,5): */ 16,
  {1,12,66,220,495,792,451,792,2,220,66,11,1}, {4,6,8,10,12,13,14,15,16,16,17,17,18,18,19,19},
  {{0,0},{1,0},{2,0},{3,0},{4,0},{5,0},{6,0},{6,1},{7,0},{8,0},{8,1},{9,0},{10,0},{11,0},{11,1},{12,0}},
  {0xF0000000,0xC0000000,0x7E000000,0x47000000,0x28100000,0x0F500000,0x08440000,0x04920000,
  0x017A0000,0x01780000,0x00818000,0x00138000,0x00030000,0x00004000,0x00002000,0x00000000} },
  { /* (24,6): */ 17,
  {1,8,65,220,2,792,924,792,495,220,59,12,1}, {4,6,7,8,9,10,11,12,13,14,15,16,16,16,17,17,17},
  {{0,0},{1,0},{1,1},{2,0},{2,1},{3,0},{4,0},{4,1},{5,0},{6,0},{7,0},{8,0},{9,0},{10,0},{10,1},{11,0},{12,0}},
  {0xF0000000,0xD0000000,0xC8000000,0x87000000,0x86800000,0x4F800000,0x4F400000,0x30700000,
  0x17B00000,0x09400000,0x03100000,0x01210000,0x00450000,0x000A0000,0x00068000,0x00008000,0x00000000} },
  { /* (24,7): */ 15,
  {1,12,66,220,495,62,924,792,495,220,66,8,1}, {5,7,9,10,11,12,13,13,14,15,15,15,15,15,16},
  {{0,0},{1,0},{2,0},{3,0},{4,0},{5,0},{5,1},{6,0},{7,0},{8,0},{9,0},{10,0},{11,0},{12,0},{11,1}},
  {0xF8000000,0xE0000000,0xBF000000,0x88000000,0x4A200000,0x46400000,0x2F700000,0x12900000,
  0x06300000,0x02520000,0x009A0000,0x00160000,0x00060000,0x00040000,0x00000000} },
  { /* (24,8): */ 15,
  {1,12,66,220,287,792,924,792,495,220,62,12,1}, {6,8,9,10,11,12,12,13,14,14,14,14,14,14,15},
  {{0,0},{1,0},{2,0},{3,0},{4,0},{4,1},{5,0},{6,0},{7,0},{8,0},{9,0},{10,0},{11,0},{12,0},{10,1}},
  {0xFC000000,0xF0000000,0xCF000000,0x98000000,0x74200000,0x67200000,0x35A00000,0x18C00000,
  0x0C600000,0x04A40000,0x01340000,0x003C0000,0x000C0000,0x00080000,0x00000000} },
  { /* (24,9): */ 14,
  {1,12,66,220,417,792,924,792,495,220,66,12,1}, {7,8,9,11,11,12,12,13,13,13,13,13,13,14},
  {{0,0},{1,0},{2,0},{3,0},{4,0},{4,1},{5,0},{6,0},{7,0},{8,0},{10,0},{11,0},{12,0},{9,0}},
  {0xFE000000,0xF2000000,0xD1000000,0xB5800000,0x81600000,0x7C800000,0x4B000000,0x2E200000,
  0x15600000,0x05E80000,0x03D80000,0x03780000,0x03700000,0x00000000} },
  { /* (24,10): */ 15,
  {1,12,66,220,221,792,923,792,495,220,66,12,1}, {7,9,10,11,11,12,12,12,12,12,12,13,13,13,13},
  {{0,0},{1,0},{2,0},{3,0},{4,0},{4,1},{5,0},{6,0},{10,0},{11,0},{12,0},{6,1},{7,0},{8,0},{9,0}},
  {0xFE000000,0xF8000000,0xE7800000,0xCC000000,0xB0600000,0x9F400000,0x6DC00000,0x34100000,
  0x2FF00000,0x2F300000,0x2F200000,0x2F180000,0x16580000,0x06E00000,0x00000000} },
  { /* (24,11): */ 14,
  {1,12,23,220,495,792,924,792,495,220,66,12,1}, {8,10,10,11,11,11,11,12,12,12,12,12,12,13},
  {{0,0},{1,0},{2,0},{2,1},{3,0},{11,0},{12,0},{4,0},{5,0},{6,0},{8,0},{9,0},{10,0},{7,0}},
  {0xFF000000,0xFC000000,0xF6400000,0xF0E00000,0xD5600000,0xD3E00000,0xD3C00000,0xB4D00000,
  0x83500000,0x49900000,0x2AA00000,0x1CE00000,0x18C00000,0x00000000} },
  { /* (24,12): */ 14,
  {1,12,66,220,495,792,504,792,495,220,66,12,1}, {10,10,10,10,11,11,12,12,12,12,12,12,12,13},
  {{0,0},{1,0},{11,0},{12,0},{2,0},{10,0},{3,0},{4,0},{5,0},{6,0},{7,0},{8,0},{9,0},{6,1}},
  {0xFFC00000,0xFCC00000,0xF9C00000,0xF9800000,0xF1400000,0xE9000000,0xDB400000,0xBC500000,
  0x8AD00000,0x6B500000,0x39D00000,0x1AE00000,0x0D200000,0x00000000} }
};

/* encoder tables (computed using decoder's tables): */
static BLADE_ENC enc_t [1+(N/2+1)+(N+1)];

/* initialize encoder: */
void blade_enc_init()
{
  unsigned int i[N+1], j, k, l, w;
  /* init enc[]: */
  for (j=0; j<1+(N/2+1)+(N+1); j++) {
    for (k=0; k<=N; k++) enc_t[j].nk[k] = dec_t[j].nk[k];
    for (k=0; k<=SGS; k++) {
      enc_t[j].sg[dec_t[j].kj[k].k][dec_t[j].kj[k].j] = j;
      enc_t[j].jen[k] = dec_t[j].jen[k];
      enc_t[j].base[k] = dec_t[j].jj_base[k] >> (32-dec_t[j].jen[k]);
    }
  }
  /* init w_ki[]: */
  for (j=0; k<=N; k++) i[k] = 0;
  for (w=0; w<(1<<N); w++) {
    for (k=0,j=0; j<N; j++) if (w & (1<<j)) k++;
    w_ki[w].k = k;
    w_ki[w].i = i[k];
    i[k] ++;
  }
}

/* initialize decoder: */
void blade_dec_init()
{
  static short b[N+1] = {1,12,66,220,495,792,924,792,495,220,66,12,1};
  unsigned int i[N+1], j, k, w;
  /* init ki_w[]: */
  for (j=0,k=0; k<=N; j+=b[k],k++) {ki_w[k] = _w + j; i[k] = 0;}
  for (w=0; w<(1<<N); w++) {
    for (k=0,j=0; j<N; j++) if (w & (1<<j)) k++;
    ki_w[k][i[k]] = w;
    i[k] ++;
  }
}

/* encoder's functions: */
unsigned blade_enc_0 (unsigned w, BITSTREAM *bs)
{
  return blade_enc (w, enc_t + 0, bs);
}

unsigned blade_enc_1 (unsigned w, unsigned cx, BITSTREAM *bs)
{
  unsigned r;
  if (cx > N/2)
    r = N - blade_enc (w ^((1<<N)-1), enc_t + 1 + N - cx, bs);
  else
    r = blade_enc (w, enc_t + 1 + cx, bs);
  return r;
}

unsigned blade_enc_2 (unsigned w, unsigned cx1, unsigned cx2, BITSTREAM *bs)
{
  unsigned cx = cx1 + cx2, r;
  if (cx > N)
    r = N - blade_enc (w ^((1<<N)-1), enc_t + 1 + (N/2 + 1) + 2*N - cx, bs);
  else
    r = blade_enc (w, enc_t + 1 + (N/2 + 1) + cx, bs);
  return r;
}

/* decoder's functions: */
unsigned blade_dec_0 (unsigned *w, BITSTREAM *bs)
{
  return blade_dec (w, dec_t + 0, bs);
}

unsigned blade_dec_1 (unsigned *w, unsigned cx, BITSTREAM *bs)
{
  unsigned b, r;
  if (cx > N/2) {
    r = N - blade_dec (&b, dec_t + 1 + N - cx, bs);
    b ^= (1<<N)-1;
  } else
    r = blade_dec (&b, dec_t + 1 + cx, bs);
  *w = b;
  return r;
}

unsigned blade_dec_2 (unsigned *w, unsigned cx1, unsigned cx2, BITSTREAM *bs)
{
  unsigned cx = cx1 + cx2, b, r;
  if (cx > N) {
    r = N - blade_dec (&b, enc_t + 1 + (N/2 + 1) + N*2 - cx, bs);
    b ^= (1<<N)-1;
  } else
    r = blade_dec (&b, enc_t + 1 + (N/2 + 1) + cx, bs);
  *w = b;
  return r;
}


/* main.c - test program and demo: */

#define M   1000    /* max # of blocks in test sequence */
#define Q   1000000 /* # of iterations */

/* test program: */
int main ()
{
  /* in/out buffers: */
  static unsigned char in_buffer [M*N/8];
  static unsigned char out_buffer [M*N/8 + 1024];
  static BITSTREAM in, out;

  /* vars: */
  unsigned char *pbs; int bit_offset;
  unsigned int w, cx, cx1 = 0, cx2 = 0;
  int i, j, k, m;
  double p, h, c;

  /* init BLADE-12 library: */
  blade_init ();

  /* scan sources: */
  for (p=0.01; p<=0.991; p+=0.01) {

    /* estimate entropy: */
    h = - (p * log(p) + (1.-p) * log(1.-p)) / log(2.);
    printf ("\np=%g, h=%g\n", p, h);

    /* try different # of blocks: */
    for (m=1; m<M; m++)
    {
      c = 0.;
      /* reset generator: */
      srand(1);
      /* make Q runs: */
      for (i=0; i<Q; i++) {

        /* generate test sequence: */
        memset(in_buffer, 0, sizeof in_buffer);
        bitstream_open(&in, in_buffer, 0, 0);
        for (j=0; j<N*m; j++) {
          /* get a next bit from a pseudo-Bernoulli source: */
          k = ((double) rand() / (double) RAND_MAX) > (1. - p);
          /* insert it in bitstream: */
          put_bits(k, 1, &in);
        }
        bitstream_close (&in, &pbs, &bit_offset, 1);

        /* start encoding: */
        memset(out_buffer, 0, sizeof out_buffer);
        bitstream_open(&out, out_buffer, 0, 0);
        bitstream_open(&in, in_buffer, 0, 1);

        /* run the encoder: */
        for (j=0; j<m; j++) {
          /* block to be encoded: */
          w = (unsigned)get_bits (N, &in);
          /* choose context and encode: */
          if (j == 0)
            cx1 = blade_enc_0 (w, &out);          /* no context */
          else if (j == 1)
            cx2 = blade_enc_1 (w, cx1, &out);     /* use cx1 */
          else {
            cx = blade_enc_2 (w, cx1, cx2, &out); /* use cx1 and cx2 */
            /* scroll contexts: */
            cx1 = cx2;
            cx2 = cx;
          }
        }
        /* close bitstreams: */
        bitstream_close (&in, &pbs, &bit_offset, 0);
        bitstream_close (&out, &pbs, &bit_offset, 1);

        /* compute coding cost: */
        c += (double)((pbs - out_buffer) * 8 + bit_offset) / (double)(m*N);

        /* start decoding: */
        bitstream_open (&in, in_buffer, 0, 1);
        bitstream_open (&out, out_buffer, 0, 1);

        /* run the decoder: */
        for (j=0; j<m; j++) {
          /* choose the context and decode: */
          if (j == 0)
            cx1 = blade_dec_0 (&w, &out);          /* no context */
          else if (j == 1)
            cx2 = blade_dec_1 (&w, cx1, &out);     /* use cx1 */
          else {
            cx = blade_dec_2 (&w, cx1, cx2, &out); /* use cx1 and cx2 */
            /* scroll contexts: */
            cx1 = cx2;
            cx2 = cx;
          }
          /* compare with the original block: */
          if (w != get_bits (N, &in)) {
            printf("?%d,", j);
          }
        }
        /* close bitstreams: */
        bitstream_close (&in, &pbs, &bit_offset, 0);
        bitstream_close (&out, &pbs, &bit_offset, 0);
      }

      /* print results: */
      c /= (double)Q;
      printf("[%d,%g], ", m*N, (c-h)/h);
      fflush(stdout);
    }
  }
  return 1;
}

\end{verbatim}
}

\end{document}